\documentstyle[12pt,epsf]{article}
\newcommand{\bi}{\bibitem}
\newcommand{\be}{\begin{eqnarray}}
\newcommand{\ee}{\end{eqnarray}}
\newcommand{\nn}{\nonumber}
\catcode`\@=11
\def\lsim{\mathrel{\mathpalette\@versim<}}
\def\gsim{\mathrel{\mathpalette\@versim>}}
\def\@versim#1#2{\vcenter{\offinterlineskip
\ialign{$\m@th#1\hfil##\hfil$\crcr#2\crcr\sim\crcr } }}
\catcode`\@=12

\begin{document}
\begin{flushright}
KANAZAWA-01-07\\
July 2001
\end{flushright}

\begin{center}
{\Large\bf Suppressing  the $\mu$ and
neutrino masses\\
by a superconformal force}
\end{center} 

\vspace{1cm}
\begin{center} {\sc Jisuke Kubo} and {\sc Daijiro Suematsu} 
\end{center}

\begin{center}
{\em 
Institute for Theoretical Physics, 
Kanazawa  University, 
Kanazawa 920-1192, Japan
}
\end{center}

\vspace{1cm}
\begin{center}
{\sc\large Abstract}
\end{center}

\noindent
The idea of 
Nelson and Strassler to obtain a power law suppression of
parameters 
by  a superconformal force is applied to understand 
the smallness of the $\mu$ parameter and neutrino masses
in R-parity violating supersymmetric standard models.
We find that the low-energy
sector  should contain at least another pair of
Higgs doublets, and that
a suppression of $\lsim O(10^{-13})$ 
for the $\mu$ parameter and
neutrino masses can be  achieved generically.
The superpotential
of the low-energy sector happens to possess an  anomaly-free discrete 
R-symmetry, either $R_3$ or $R_6$,
which naturally suppresses certain lepton-flavor violating
processes, the neutrinoless double beta decays and
also the electron electric dipole moment.
We expect
that the escape energy of the superconformal sector is 
$\lsim$ O(10) TeV so that this sector will be observable at LHC.
Our models can accommodate to
a large mixing among neutrinos and give
the same  upper bound
of the lightest Higgs mass  
as the minimal supersymmetric standard model.

\vspace{1cm}
\noindent
PACS number:12.60.Jv, 11.30.Fs, 14.60.Pq

\newpage
\section{Introduction}
 The minimal  supersymmetric standard model (MSSM)
contains two Higgs chiral supermultiplets,   $H_u$  and $H_d$, and
with respect to the standard model (SM)
gauge group $SU(3)_C\times SU(2)_{L}\times U(1)_Y$  the down-type Higgs
doublet $H_d$  has the same
quantum numbers as the  left-handed lepton doublets $L_i ~ (i=1,2,3)$.
Therefore, the $SU(3)_C\times SU(2)_L\times U(1)_Y$ gauge interactions cannot
distinguish $H_d$  from $L_i$. What distinguish them from each other
are  lepton number and  R-parity \cite{farrar},
which, however, forbid Majorana neutrino masses.
An elegant way to generate  small neutrino masses is the
 see-saw mechanism \cite{yanagida}, and if we apply this mechanism
without  breaking R-parity we have to introduce right-handed
neutrinos into the MSSM. It has been  known for long time
that once we give up the lepton number  as well as R-parity conservations,
there exist  possibilities to generate neutrino masses
through mixing with neutralinos
without  introducing the  right-handed neutrinos  
\cite{farrar}, \cite{rmass1}--\cite{suematsu1}.

In this paper we are concerned with these R-parity violating 
(RPV) models \footnote{See \cite{report} for recent developments.}.
In the RPV models, there exists no difference among   $H_d$  and $L_i$.
That is, the $\mu$ term,  $H_u H_d$,
and the  bilinear RPV terms,  $H_u L_i$,
should be treated on the same footing, which implies that
the $\mu$ problem  \cite{kim} 
\footnote{See also \cite{nmssm}--\cite{musol} for various possible
solutions for the $\mu$-problem.} is
 closely related to the smallness of 
the  neutrino masses \cite{murayama}. 
So, unless the $\mu$ problem is solved, the natural neutrino
mass in the RPV models will be of the order of a fundamental scale,
which is a disaster for the models.
Our basic idea to obtain a small $\mu$ and hence
small neutrino masses is to use a superconformal
strong force  to drive $\mu$ down to the electroweak scale from a superhigh 
energy scale. A similar idea has been applied in the Yukawa sector 
and in the supersymmetry breaking sector by Nelson and Strassler 
\cite{nelson} 
\footnote{The basic  mechanism will be explained in the text, and
for more details see \cite{nelson}.
See also \cite{kobayashi1,nakano}. } to
generate a hierarchical order of the Yukawa couplings at low energies 
from their  anarchical order at a fundamental scale, and at the same time
to obtain almost degenerate 
soft-supersymmetry-breaking (SSB) scalar masses 
at low energies  \cite{karch,luty} .

For our idea to work, we have to couple the Higgs fields to 
a superconformal sector. However, if the MSSM Higgs multiples
couple to the strong sector, not only $\mu$, but also all the
Yukawa couplings are suppressed, which we would like to 
avoid in this paper. So, we will enlarge the Higgs sector.
We introduce another pair of Higgs doublets,
$\tilde{H}_{u}$ and $\tilde{H}_{d}$, which are supposed 
to couple to the superconformal
sector and are responsible to drive  $\mu$
down  to the electroweak scale.
We will find that a suppression of $\lsim O(10^{-13})$ can be achieved
in this way, and we expect that
the escape energy of the superconformal sector is 
 rather a lower scale $\sim O(\mbox{TeV})$, because
otherwise  the superconformal
suppression would be insufficient to
understand the smallness of the $\mu$ and  neutrino masses.
Since the charged matters  in the superconformal sector
have nontrivial quantum numbers under
$SU(2)_L\times U(1)_Y$, they could be 
experimentally tested at LHC, for instance.

We will explicitly construct realistic low-energy models 
by imposing anomaly-free discrete R-symmetries \cite{ibanez} 
in the superpotential,
while allowing most general, renormalizable  supersymmetry breaking
terms. It will turn out
that our models can accommodate to
a large mixing among neutrinos, and
that  the  upper bound
of the lightest Higgs mass  
of the MSSM remains unchanged.

\section{Superconformal Sector}   

We assume that the superconformal gauge force that suppresses  $\mu$ 
is based on the gauge group $SU(N_C)$
with a global symmetry 
$U(N_{TS})_L\times U(N_{TS})_R\times U(N_U)_L\times U(N_U)_R$.
The  matter content is given in Table 1.
\begin{figure}[hbt]
\begin{center}
\begin{tabular}{|c|c|c|c|c|}
\hline
 & $SU(3)_C\times SU(2)_{L}\times U(1)_Y$ 
& $SU(N_C)$ & $U(N_{TS})_L\times U(N_{TS})_R$
& $ U(N_U)_L\times U(N_U)_R$  
\\ \hline
$T$ &$ (1,2,-1/2)$ & $N_C$ &
$(N_{TS},1)$  & $1$ 
\\ \hline
$\overline{T}$ &$ (1,2,1/2)$ & $\overline{N}_C$ &
$(1,\overline{N}_{TS})$  & $1$
\\ \hline
$S$ &$ (1,1,0)$ & $N_C$ &
$(1,N_{TS})$  & $1$
\\ \hline
$\overline{S}$ &$ (1,1,0)$ & $\overline{N}_C$ &
$(\overline{N}_{TS},1)$  & $1$ 
\\ \hline
$U$ &$ (1,1,0)$ & $N_C$ &
$1$  & $(N_{U},1)$ 
\\ \hline
$\overline{U}$ &$ (1,1,0)$ & $\overline{N}_C$ &
$1$  & $(1,\overline{N}_{U})$ 
\\ \hline
\end{tabular}
\vspace*{1mm}

\footnotesize
{\bf Table 1}. Field contents in the superconformal sector. 
\end{center}
\end{figure}
\normalsize
Note that the representations of the matter chiral supermultiplets in 
this sector should be real with respect to the SM gauge symmetry
$SU(3)_C\times SU(2)_{L}\times U(1)_Y$.
Otherwise the strong force could break dynamically these symmetry,
at least  at the escaping energy scale $\Lambda_{C}$, at which the 
strong sector is
supposed to decouple from the low-energy sector.
(We will estimate $\Lambda_{C}$ later on.) This implies that
the representation of the new Higgs supermultiplets 
$\tilde{H}_{u}$ and $\tilde{H}_{d}$ that couple to the superconformal sector
should also be  real with respect to these symmetries.
With this remark we now consider the coupling
of $\tilde{H}_{u}$ and $\tilde{H}_{d}$ to this
sector through the renormalizable superpotential
\be
W_{SC} &=&
y_{U}\tilde{H}_{u} T \overline{S} +y_{D} \tilde{H}_{d} 
\overline{T} S~,
\label{wsc}
\ee
where we have suppressed all the indices, and
the new Higgs doublets $\tilde{H}_{u}$ and $\tilde{H}_{d}$
are singlets under $SU(N_C)$, where  the $U(1)_Y$ charge of
$\tilde{H}_{u(d)}$ is $+(-)1/2$.

Let us shortly explain the mechanism proposed by 
Nelson and Strassler \cite{nelson}
using our model.
According to Seiberg's conjecture \cite{seiberg},
 a nontrivial infrared fixed 
point  exits in our model\footnote{
We assume that the supersymmetric SM sector (SSM)
couples only weakly to the strong sector and so
 the conjecture is approximately satisfied.},
if $(3/2)N_{C} < 3 N_{TS}+N_U <3 N_C$ is satisfied \cite{seiberg}.
The anomalous dimension $\gamma_{I}$ of 
a chiral supermultiplet
$\phi_{I}$
at the fixed point is related to its charge $R_I$
of an anomaly-free R-symmetry  through
$\gamma_I = (3/2)R_I -1$ \cite{seiberg,flato}.
(We assume below that $\overline{T}$,
$\overline{S}$ and $\overline{U}$ have, respectively,
the same anomalous dimensions as
$T$, $S$ and $U$.)
The  point is that the anomalous dimensions
 can become  large
negative numbers,
because the contribution of  gauginos with
a positive R charge to the anomaly has to be
cancelled by that of chiral charged matter supermultiplets
with negative R charge.
This can also be seen from the 
Novikov-Shifman-Vainstein-Zakharov
$\beta$-function \cite{novikov1}
 \be
 \beta(g) & =&-{g^3\over 16\pi^2}{3N_C-
 3  N_{TS}-N_U +2\Gamma
 \over   
 1-N_C g^2/8\pi^2}~,\quad
 \Gamma=N_{TS} (2\gamma_{T}+\gamma_{S})+N_U\gamma_U~.
 \label{beta}
 \ee
So, at the fixed point we obtain
 \be
\Gamma  &=&-\frac{1}{2}( 3N_C-(3N_{TS}+N_U))
~,  \quad
(\frac{3}{2}N_C < 3N_{TS}+N_{U} < 3 N_C)~.
 \ee
If we may assume that all the chiral 
 supermultiplets have the same anomalous dimension 
 $\gamma$ for simplicity, we find that
 \be
 \gamma &=& -\frac{3N_C-(3N_{TS}+N_U)}{2(3N_{TS}+N_U)}
 \ee
 at the fixed point, implying that the anomalous dimensions
 can become  negative numbers of $O(1)$. Further,
 at the superconformal fixed point, the  dimension of the 
superpotential $W_{SC}$ 
 has to be  $3$ which means that its anomalous dimension
 should vanish. Therefore, we arrive at
 \be
\gamma{*} &=&  \gamma_{\tilde{H}_{u}} =  \gamma_{\tilde{H}_{d}}
 =-2\gamma=  \frac{3N_C-(3N_{TS}+N_U)}{3N_{TS}+N_U} < 1~,
 \label{gammastar}
 \ee
 which is a positive number of $O(1)$, and for instance,
$1/14\leq \gamma^{*} \leq 7/8 $ for $SU(5)$.

The crucial point is now that the large positive anomalous dimension
 $\gamma^{*}$ carried by the SSM supermultiplets has 
 a large influence on the SSM parameters if their evolution
 has the form
\be
\Lambda \frac{d \mu}{d \Lambda}  &= & \mu ~
\gamma_{\tilde{H}_{u,d}}+\cdots~,
\ee
where $\cdots$ stand for other contributions from the SSM,
which are assumed to be small at high energies.
If the energy goes down from a unification scale $\Lambda_0$ (which may be the 
Planck scale, string scale or GUT scale) to the escaping scale
 $\Lambda_C$ at which the 
strong sector decouples due to some dynamics, the parameter $\mu$ enjoys
the strong suppression of the form
 \be
 \mu(\Lambda_{C}) &\simeq  \mu(\Lambda_{0}) 
 ~ \left[\Lambda_{C}/\Lambda_{0}\right]^{\gamma^{*}}~.
 \label{sup0}
 \ee
This is the mechanism of  suppression \cite{nelson}, and we assume that
all the massive supersymmetric parameters in the superpotential of the
SSM sector enjoy this suppression.
 
Note, however, that the anomalous dimension at the
superconformal fixed point  cannot exceed $1$, 
if only one chiral multiplet couples to the parameter.
That is, 
\be
\frac{ \mu(\Lambda_{C})}{ \mu(\Lambda_{0})} & > &
\frac{ \Lambda_{C}}{ \Lambda_{0}}~,
\label{sup1}
\ee
so that if we would identify $ \mu(\Lambda_{0})$ with $
 \Lambda_{0}$, we would obtain a  useless result
  $\mu(\Lambda_{C}) > \Lambda_{C}$.
  A consequence of this observation is that above the 
  unification scale $\Lambda_0$
  the parameter $\mu$ should have  already  enjoyed
  some suppression mechanism which yields a suppression of
 \be
 \frac{\mu(\Lambda_0)}{\Lambda_0} &\simeq & 
 \frac{\mu(\Lambda_C)}{\Lambda_C}
 \left [\frac{\mu(\Lambda_C)}{\mu(\Lambda_0)}\right]^{1/\gamma^*-1}~.
 \label{sup2}
 \ee
 The value of  $\gamma^*$ is typically $\lsim 0.8$. Assuming
 that $1/\gamma^*-1 \simeq 0.2$, $ \Lambda_C/\mu(\Lambda_C)\simeq 50$ and
 $\mu(\Lambda_C)/\mu(\Lambda_0)\simeq 10^{-10}$,
 we obtain a necessary suppression of
 $ \mu(\Lambda_0)/\Lambda_0 \simeq 10^{-4}$.

 Before we come to construct the SSM sector, let us 
 compute the anomalous dimensions $\gamma^*$ in our model
 in a semi-nonperturbative way.
 That is, we use the nonperturbative result for the $\beta$-function
 of the gauge coupling ($\ref{beta}$), but for the anomalous dimensions we 
 use the one-loop expression
 \be
 \gamma_{\tilde H_u} &=&\frac{1}{16 \pi^2}N_{TS}~y_U^2 ~, \quad
  \gamma_{T} =\frac{1}{16 \pi^2}(~y_U^2-\frac{N_C^2-1}{N_C}g^2) ~,
  \label{gamma1} \\
   \gamma_{\overline{S}} &=&
\frac{1}{16 \pi^2}(~2 y_U^2-\frac{N_C^2-1}{N_C}g^2) ~, \quad
  \gamma_{U} =
-\frac{1}{16\pi^2}~\frac{N_C^2-1}{N_C}g^2~,
  \label{gamma2}
 \ee
 and similarly for $ \gamma_{\tilde H_d}$ etc..
 From $\beta(g)=0$ and $ \gamma_{\tilde H_u}+\gamma_{T}
+\gamma_{\overline{S}}
 = \gamma_{\tilde H_d}+\gamma_{\overline{T}}+\gamma_{S}=0$,
 we obtain
 \be
\gamma^* &=&  \gamma_{H_u}=\gamma_{H_d}=
N_{TS}\frac{3(N_C-N_{TS})-N_U}{N_{TS}+3 N_{TS}^2+3N_U+N_{TS}N_U}~.
\label{gamma3}
 \ee
The maximal value  $\gamma^{*}_{max}$ for
a given gauge group can be computed from Eq. (\ref{gamma3}). 
We find for instance
\be
\gamma^{*}_{max}(SU(3)) &=&\frac{1}{3}
~, \quad \gamma^{*}_{max}(SU(5)) =  \frac{7}{12} ~,\nn\\
\gamma^{*}_{max}(SU(7)) &= &  \frac{5}{7} ~, \quad
\gamma^{*}_{max}(SU(9)) =  \frac{26}{33}~.
\ee
Note that the numbers above are not exact results,
because we have used only one-loop anomalous 
dimensions in Eqs. (\ref{gamma1}) and (\ref{gamma2}).
(In some cases,  one-loop anomalous dimensions yield exact results.)
So these numbers may receive nonperturbative corrections.

As we have seen in  this section, the superconformal force can
suppress $\mu$ according to the power law (\ref{sup0}).
However, the suppression 
$\mu(\Lambda_C)/\mu(\Lambda_0)$ is not strong enough so that 
only a suppression of $\gsim 0(10^{-13})$ 
can be gained from the superconformal force
if we assume that
$\Lambda_C/\Lambda_0 \gsim 10^{-16}$ and $\gamma^* \lsim 0.8$,
where we have used (\ref{sup2}).
We therefore cannot identify $\mu(\Lambda_0)$ with the fundamental scale
$\Lambda_0$ so that we have to assume that
a suppression of at least $\lsim 10^{-3}$
should  already  exist in the fundamental theory.
A representative example is:
\be
\Lambda_C &\simeq &  1.8 ~\mbox{TeV}~,~
\mu(\Lambda_C) \simeq  10^2 ~\mbox{GeV}~,
~\mu(\Lambda_0) \simeq  10^{13} ~\mbox{GeV}~,~
\Lambda_0  \simeq   ~10^{17}~\mbox{GeV}~,
\ee
where we have assumed that $\gamma^*=0.8$.
This should be contrasted to
the models of Nelson and Strassler \cite{nelson},
where $\Lambda_C$ is supposed to be between
$10^{10}$ and $10^{16}$ GeV.
Our models predict rather lower scale $\sim O(\mbox{TeV})$, because
otherwise  the superconformal
suppression would be insufficient to
understand the smallness of $\mu$ and  neutrino masses.
Since the charged matter multiplets  in the superconformal sector
have nontrivial quantum numbers under
$SU(2)_L\times U(1)_Y$, they could be produced
and seen as new type of hadrons at LHC.

 \section{The low-energy sector}
 We assume that the low-energy physics can be described by a 
 supersymmetric extension of the SM and that all the 
 supersymmetric mass parameters enjoy the superconformal
 suppression. 
 As explained in the introduction, we have to enlarge the matter content of the
 MSSM for this idea to work, and we have already introduced,
 in addition to the MSSM Higgs doublets
 $H_{u}$ and $H_{d}$,
a new set of Higgs doublets $\tilde{H}_{u}$ and $\tilde{H}_{d}$
 which couple to the superconformal sector.
 The SM gauge interactions cannot distinguish
  $\tilde{H}_{u}$ from $H_{u}$  and 
  $\tilde{H}_{d}$ from $H_{d}$, and so we would like to
  find a symmetry which makes it possible
  to distinguish them from each other and allows in the superpotential
  the quadratic terms  like  $H_{d}\tilde{H}_{u}$, 
  $H_{u}\tilde{H}_{d}$,  but forbids
  $H_d H_u$ (which has to be absent,
  because it cannot enjoy the superconformal suppression).
 First we consider an ordinary global $U(1)$ 
  or discrete $Z_N$ symmetry\footnote{By an ``ordinary'' symmetry 
we mean a symmetry which is not of  R-symmetry type.},
  and we assume that  the superconformal
  strong force does not break
  nonperturbatively the symmetry. This implies that the representations
  of the charged matter multiplets in the strong sector should be 
  real with respect
  to  the  symmetry, that is,
   the $U(1)$ (or $Z_N$) charge of $\tilde{H}_{u}$  
  has to be the opposite sign of that of $\tilde{H}_{d}$.
 Consequently,  
  $\tilde{H}_{u}\tilde{H}_{d}$  and hence 
 $H_{u}H_{d}$ cannot be forbidden 
 by an ordinary global $U(1)$ 
  or discrete $Z_N$ symmetry if 
    $H_{d} \tilde{H}_{u}$ and 
  $H_{u}\tilde{H}_{d}$ are allowed.
 
Another possibility is R-symmetry, discrete or continuos.
We understand under the reality of a R-symmetry in the strong
sector that 
the charged matter multiplets, $T, S$ and $U$ can form
a mass term with $\overline{T}, \overline{S}$ and 
$\overline{U}$, respectively.  So, their R-charge has to be one,
implying that the charge of $\tilde{H}_{u}$  and
$\tilde{H}_{d}$  has to be zero such that the Yukawa coupling
(\ref{wsc}) is allowed by the symmetry. 
We look for an anomaly-free   R-symmetry along the line of 
  \cite{ibanez,kurosawa}, because  such a symmetry may descend
  from a gauge symmetry in a compactified string theory.
We denote the R charge of a chiral supermultiplet $\phi$
  by $R(\phi)$, and impose  the following conditions:
  \begin{enumerate}
      \item
      The reality of  $( \tilde{H}_{u}~,~\tilde{H}_{d})$,
      which means $R(\tilde{H}_{u})=R(\tilde{H}_{d})=0$.
      
      \item
      The presence of  $H_{d} \tilde{H}_{u}$ and   $H_{u}\tilde{H}_{d}$.
    
      \item 
      The absence of $H_{u}H_{d}$.
      \item 
      The presence of the Yukawa terms $E^c_i L_j H_{d}$, $D^c_i Q_j H_{d}$
      and $U^c_i Q_j H_{u}$.

     \end{enumerate}   
Here   $E_i$, $U_i$ and $D_i$ are the right-handed lepton,
up-type quark and down-type quark singlets,  and
  $L_i$ and $Q_i$ are the left-handed lepton,
quark doublets ($i=1,2,3$), respectively. 
An immediate 
consequence of the reality condition 1 is
that  $R_2$ (R-parity) is ruled out, because this condition 
implies that
$R(H_{d})=R(H_u)=2=0\ ({\rm mod}~2)$ due to the condition 2, which however
contradicts with the condition 3. So we will not consider $R_2$
in the following discussion.
The conditions 1 and 2 
yield that
\be
R(H_{d}) &=& -R(\tilde{H}_{u})+2
=2 \quad (\mbox{mod}~N)~,~\nn\\
R(H_{u}) &=&  -R(\tilde{H}_{d})+2=
2\quad (\mbox{mod}~N)~,
\label{cond2}
\ee
 which give
 \be
 R(H_{u})+R(H_{d}) &=&4\quad (\mbox{mod}~N)~,
 \label{cond3}
 \ee
where we have take into account the possibility that the R-symmetry
may be a discrete symmetry $R_N$.
The last condition 4
requires
\be
&&R(H_{u})+R(Q_i)+R(U^c_j)=R(H_{d})+R(Q_i)+R(D^c_j) \nn\\ 
&&\hspace*{1cm}=R(H_{d})+R(L_i)+R(E^c_j)=2
\quad ({\rm mod}~ N).
 \label{cond4}
\ee
One can easily see that Eq. 
 (\ref{cond4}) requires that the trilinear terms
\be
D_i^c Q_j  \tilde{H}_{d}~~\mbox{and}~~U_i^c Q_j  \tilde{H}_{u}
\ee
should be absent. 

There exist mixed non-abelian 
gauge anomalies, $R [U(1)_Y]^2$,
$R [SU(2)_L]^2$, $R [SU(3)_C]^2$
and $R [SU(N_C)]^2$, the cubic $R^3$ and  mixed
gravitational anomalies. The cubic and  mixed
gravitational anomalies depend on the 
structure of the massive states
in  the high-energy theory (so they do not decouple
in a certain sense at low-energies \cite{ibanez}), while
the mixed gauge anomalies should be cancelled
by the massless fermions \cite{ibanez,banks}.
Since we are not interested in the high-energy sector
in the present paper, we would like to take into account
only the mixed gauge anomalies.
Moreover, the $R [U(1)_Y]^2$ anomaly does not give 
useful information, because the $U(1)_Y$ charge is not
quantized. With these remarks in mind, we  consider
$R [SU(2)_L]^2$ and $R [SU(3)_C]^2$ only.
The anomaly coefficients are given by \cite{ibanez,banks,kurosawa}
\be
{\cal A}_2 &=&
\frac{3}{2}~3~\Big( R(Q)-1 \Big)+\frac{1}{2}\sum_{i=1}^{3}
\Big( R(L_i)-1 \Big)
+\frac{1}{2}\Big[~\Big( R(H_{u})-1 \Big)+\Big( R(H_{d})
-1 \Big)   \nn\\
& &+\Big( R(\tilde{H}_{u})-1 \Big)+
\Big( R(\tilde{H}_{d})-1 \Big)  ~\Big]+2~,
\label{ano2}\\
{\cal A}_3 &=&
\frac{3}{2}\Big[~2 \Big(R(Q)-1\Big)+\Big(R(U)-1\Big)
+\Big(R(D)-1\Big)~\Big]+3~,
\label{ano3}
\ee
where we have considered the possibility that
the R charge of the leptons may depend on the generation,
while  we have assumed that for quarks
it is independent of the generation.
Using now Eqs. (\ref{cond2}) -- (\ref{cond4}), 
the anomaly coefficients  (\ref{ano2}) and (\ref{ano3})
can be rewritten as
\be
2 {\cal A}_2 &=&\Big[~-8+\sum_{i=1}^{3}\Big(R(L_i)+9
R(Q)\Big)~\Big]\quad  (\mbox{mod}~N)~,
\label{anomaly2}\\
2 {\cal A}_3 &=&6 ~\Big[~1-\frac{1}{2}\Big(R(H_{u})+R(H_{d}) 
\Big)~\Big] = -6 \quad  (\mbox{mod}~N)~.
\label{anomaly3}
\ee
Eq. (\ref{anomaly3}) implies that a continuos R-symmetry cannot be anomaly free.
So we look for anomaly-free discrete R-symmetries $R_N$.
For $R_N$, the right-hand side of 
Eqs. (\ref{anomaly2}) and (\ref{anomaly3}) may  be
$Nk$ to ensure anomaly-freedom, where $k$ is an 
arbitrary integer. Therefore, Eq. (\ref{anomaly3}) 
implies that we can have only $R_3$ or
$R_6$ ($R_2$ has already been ruled out).
Another immediate consequence is that if $R(L_{i})$ is independent
of the generation, $2{\cal A}_2=Nk$ cannot be satisfied
for $N=3$ and $6$, because $8$ cannot be cancelled by
a multiple of three. In the following discussion we will
assume that $L_1$ has a  $R$ charge  that is different from those of 
$L_2$ and $L_3$ (although there are other possibilities, e.g.
that the R charge of the quarks is generation dependent).
The $R [SU(N_C)]^2$ anomaly results only from the 
 $SU(N_C)$ gauginos (the condition 1 is a consequence
 of $ R(T)=R(\overline{T})=\cdots=R(\overline{U})=1$ ):
 \be
2{\cal A}_{N_C} = 2T\Big(SU(N_C)\Big)= ~2 N_C~,
\ee
 which implies that,
 because of $R_3$ or $R_6$, only  a multiple of $3$  for $N_C$ is possible.

We have checked that there exist various solutions, and we would like 
to give here only two  representative solutions in Table 2.
The models also possess the Baryon triality symmetry $B_3$ \cite{ibanez}
which is free not only from the mixed non-abelian gauge
anomalies, but also from the cubic as well as the mixed gravitational
anomalies \footnote{
The Baryon triality is defined as 
$B_3=2Y-B~({\rm mod}~3)$ \cite{ibanez},
where $Y$ and $B$ are the weak hypercharge and Baryon number, respectively.
The Baryon triality assignment in the superconformal sector
is not unique. A possibility is that $B_3(T)=2,B_3(\overline{T})=1$
and all the other superfields have zero charge.}.

\begin{figure}[hbt]
\begin{center}
\begin{tabular}{|c|c|c|c|c|c|c|c|c|c|c|c|}
\hline
$R$ &
$H_{u}$ & $H_{d}$ &
$\tilde{H}_{u}$  & $\tilde{H}_{d}$ &
$L_{1}$ & $ L_{2,3}$ & $ E_{1}^c$ & $E_{2,3}^c$ &
$Q$ & $U^c$ & $D^c$ \\ \hline
$R_3$ &
$2$ & $2$ & $0$  & $0$ & $1$ & $2$ & $2$ & $1$ & $0$ & $0$ & $0$  \\ \hline
$R_6$ &
$2$ & $2$ & $0$  & $0$ & $4$ & $2$ & $2$ & $4$ & $0$ & $0$ & $0$  \\ \hline
$B_3$ & $1$ & $2$ & $1$  & $2$ &$2$ & $ 2$ & $ 2$ & $2$ & $0$ & $2$ &
 $1$  \\ \hline
\end{tabular}
\end{center}

\footnotesize
{\bf Table 2}. The R charge assignment of two representative models.
The last row is the Baryon triality \cite{ibanez}.
\end{figure}

\normalsize
\vspace{0.2cm}

\noindent
The superpotential corresponding to the $R_3$  and $R_6$ models
takes the form
\be
W &=& W_\mu+W_Y+W_{Y}'~,
\label{superp}
\ee
where
\be
W_\mu &=& 
\tilde{\mu} H_{u}\tilde{H}_{d}+\mu_0 H_{d}\tilde{H}_{u}+\sum_{i=2,3}
\mu_i L_i\tilde{H}_{u}~,
\label{superp2}\\
W_Y &=&\sum_{i,j=2,3}~y_{i j}^e~ L_i H_{d}E_j^c + 
y_{11}^e~ L_1 H_{d}E_1^c
+ \sum_{i,j=1}^3 \left[y_{ij}^d~ Q_i H_{d} D_j^c
+y_{i j}^u~Q_i H_{u}U_j^c\right] ~, \nn\\
W_{Y}' &=&
\sum_{i=2,3}\Big[~ \lambda_{i11}L_1 L_i E_1^c~+\lambda_{23i}L_2 L_3 E_i^c
+\sum_{j,k=1}^3(~\lambda^{\prime}_{ijk} L_i Q_j D^c_k
+\tilde{y}_{1i}^eL_1\tilde{H}_{d}E_i^c ~)~\Big]~.
\label{superp3}
\ee
The coupling constant $\lambda_{ijk}$ is antisymmetric with respect to
the first two indices ($\lambda_{ijk}=-\lambda_{jik}$). 
The last term $\tilde{y}_{1i}^e~L_1\tilde{H}_{d}E_i^c$ in $W_Y^\prime$ 
could cause a FCNC problem, but it is not, because
$\tilde{y}_{1i}^e$ will be extremely suppressed by
the superconformal force.
Note that the baryon number violating term $D^c D^c U^c$
is absent in the superpotential.
This term is protected by $B_3$ and also by the discrete R-symmetry.

To make our model viable we have to take into account supersymmetry
breaking. We assume that it appears  as 
 soft-supersymmetry-breaking  (SSB) lagrangian ${\cal L}_{soft}$.
What about symmetry of ${\cal L}_{soft}$?
 If we impose the same global symmetry $R_3$ or $R_6$ 
 on ${\cal L}_{soft}$, the gaugino mass
 terms  for instance are not allowed. This would be phenomenologically 
a  disaster.
In the case of the MSSM,
the SSB terms satisfy $R_2$ symmetry (R-parity),
and moreover this symmetry is  free of
all anomalies.
But
the superpotential of the MSSM with or without
RPV terms has  a lager R-symmetry than $R_2$
which is free from mixed non-abelian gauge anomalies.
One can convince oneself for instance
that  an anomaly-free $R_4$ or  $R_5$  is
realized
in the superpotential.
These discrete symmetries $R_4$ and $R_5$ are assumed be completely broken 
by the SSB terms in the case of the MSSM,
while the completely anomaly-free $R_2$ is unbroken
by the SSB terms.
In the present case we therefore assume that 
the completely anomaly-free $B_3$ is unbroken, while 
the superpotential symmetry, $R_3$ or $R_6$, 
is broken by the SSB terms. We thus include
all renormalizable SSB terms in ${\cal L}_{soft}$
that are consistent with $B_3$.
Then the SSB Lagragian 
is given by
\be
-{\cal L}_{soft}& =&
\sum_{i,j=1}^2(\tilde{m}^2_u)_{ij}H^{*}_{ui} H_{uj}
+\sum_{\alpha,\beta=1}^5(\tilde{m}^2_d)_{\alpha\beta}H^{*}_{d\alpha}H_{d\beta}
\nn \\
& &+\sum_{i,j=1}^3\Big[(\tilde{m}^2_Q)_{ij}Q^{*}_{i} Q_{j}
+(\tilde{m}^2_U)_{ij}U^{*c}_{i}U_{j}^c 
+(\tilde{m}^2_D)_{ij}D^{*c}_{i}D_{j}^c\Big] \nn\\
& &+\left[~-\sum_{i=1}^2\sum_{\alpha=1}^5B_{i\alpha} H_{ui}H_{d\alpha}
+\sum_{i=1}^3\sum_{\alpha,\beta=1}^5h_{\alpha\beta i}^e
H_{d\alpha}H_{d\beta} E_i^c \right.\nn   \\
& &\left.+
\sum_{i,j=1}^3\left(\sum_{\alpha=1}^5h_{i j\alpha}^d  Q_{i} H_{d\alpha} D_j^c
 +\sum_{k=1}^2h_{i j k}^u  Q_{i} H_{uk}U_j^c\right) +h.c.~\right]~,
\label{soft1}
\ee
where the gaugino masses are abbreviated and
the same notation has been used for the scalar component
of a  supermultiplet as the corresponding superfield. 
We have denoted the Higgs doublets  $H_u$ and $\tilde{H}_{u}$ by
$ H_{ui}$ with $i=1,2$, and the down-type ones
 $H_d, \tilde{H}_{d}$ and $L_i~(i=1,2,3)$
 by $ H_{d\alpha}$ with $\alpha=1,\dots,5$, respectively. 

The superpotential (\ref{superp3}) has various 
phenomenological  consequences.
First of all there is no Baryon decay as emphasized.
($\lambda_{ijk}^{\prime\prime}$ in the notation
of \cite{report} vanish identically.)
Further various Yukawa couplings vanish:
\be
&& y_{1i} = \lambda_{231}=\lambda_{1ij}=0  \quad \mbox{for}~ i,j=2,3~,~ \nn\\
&& \lambda_{1ij}^{\prime}=0 \quad \mbox{for}~ i,j=1,2,3.
\label{yukawas}
\ee
Therefore, the bounds coming from a certain set of the lepton-flavor
violating processes such as $\mu\rightarrow e~\gamma,
~\mu\rightarrow e~e~e,$ $\mu$-$e$ conversion in nuclei
\cite{choudhury}-\cite{gouvea}, the electron EDM\cite{frank} and 
the neutrinoless double $\beta$ decay \cite{doi,hirsch,mohapatra} 
are automatically satisfied.
But the lepton-flavor violating $\tau$ decays as well as
various RPV rare leptonic decays of light mesons\cite{choudhury}
 such as
 $K_L\rightarrow \mu\bar\mu,
~K_L\rightarrow e \bar e$ are allowed, while
 a certain mode such as $K_L\rightarrow e\bar\mu+\bar e\mu$ 
is forbidden, giving
 constraints on the
RPV Yukawa couplings\footnote{
It is assumed here and above that the mass of all the scalar quarks and
leptons is 100 GeV.}\cite{choudhury,report}
\be
\lambda_{232}\lambda_{312,321}^{\prime} ~{^<_\sim}~ 3.8\times 10^{-7}~,\qquad
\lambda_{121}\lambda_{212,221}^{\prime}~,~\lambda_{131}\lambda_{312,321}^{\prime}
~{^<_\sim}~ 2.5\times 10^{-8}~.
\ee
These might be considered as prediction of the present model and
make it possible to discriminate the model from other RPV models.
There are other phenomenological consequences, which we
would like to leave for future work.

 \section{Neutrino mass and the lightest Higgs mass}
\subsection{Neutrino mass and mixing}
First we would like to derive 
the neutralino-neutrino mass matrix $M$
for the superpotential (\ref{superp}) along with
the SSB lagrangian (\ref{soft1}).
To this end, we define the neutralino vector as
\be
\Psi^T &= &(-i\lambda_1,-i\lambda_2,\psi_u, \psi_{\tilde{u}}
~\psi_d,\psi_{\tilde{d}},\psi_i)~,~i=1,2,3~,
\ee
where $\lambda_{1,2}$ are the gauginos for  $U(1)_Y$ and 
$SU(2)_L$, and $\psi$'s are the neutral fermionic components
of the Higgs and left-handed lepton supermultiplets
in an obvious notation.
The vacuum expectation values (VEVs) of 
the neutral bosonic components
of the Higgs and left-handed lepton supermultiplets are denoted
by $v_{I}$ with $I=u,\tilde{u}, \dots$, and
 our normalization of $v$'s can be read off from
 \be
 v & =& \frac{2 M_W}{g}=246~\mbox{GeV}~,\qquad v^2=\sum_{I=u,\tilde{u},\dots}
 ~v_{I}^2~,
 \ee
 where $g$ is the $SU(2)_L$ gauge coupling constant,
 and $M_W$ is the $W$ gauge boson mass.
We also use the notation $v_0=v_{d}$, 
 $\rho_{I}=v_{I}/v$,
 $M_{sw} = M_Z~\sin\theta_{W}=
 M_W \tan\theta_W$ and $M_{cw}= M_Z~\cos\theta_{W}=M_W$,
 where $\theta$ is the Weinberg angle.
Then neutralino-neutrino mass term can be written as
$-(1/2)\Psi^T~M~\Psi$, where
\be 
M&=&\left(\begin{array}{cc} {\cal M}_0 & {\cal M}^T\\ 
{\cal M} & 0 \\\end{array} \right)~, 
\ee
\be
{\cal M}_0&=&\left(
\begin{array}{cccccc} 
    M_1 &0 &  M_{sw} \rho_u &  M_{sw} \rho_{\tilde{u}} & 
    -M_{sw} \rho_0  & -M_{sw} \rho_{\tilde{d}} \\
0 &M_2 & -M_{cw} \rho_u &-M_{cw} \rho_{\tilde{u}} & 
    M_{cw} \rho_0  &  M_{cw} \rho_{\tilde{d}} \\
M_{sw} \rho_u &-M_{cw} \rho_u & 0 &0 & 0  & \tilde{\mu} \\
 M_{sw} \rho_{\tilde{u}} &-M_{cw} \rho_{\tilde{d}}  &0 & 0& \mu_0 &0 \\
-M_{sw} \rho_0 & M_{cw} \rho_0  & 0 &\mu_0 & 0 &0 \\
- M_{sw} \rho_{\tilde{d}} & M_{cw} \rho_{\tilde{d}}  &
\tilde{\mu} &0 &  0  &0 \\
  \end{array} \right)~, \nn \\
{\cal M}&=&\left(\begin{array}{cccccc}
 -M_{sw} \rho_i & M_{cw} \rho_i  &0 &\mu_i & 0  &0 \\
\end{array}\right)~.
\ee
Here ${\cal M}_0$ is a neutralino mass matrix and a neutralino-neutrino mixing 
matrix is represented by  ${\cal M}$.
Through this neutralino-neutrino mixing neutrinos can be massive as
discussed in the usual RPV models \cite{rmass1}--\cite{suematsu1}.

The smallness of the neutrino masses can be achieved in two ways. 
One possibility is given  by a precise alignment of $\vec{\rho}$ 
and $\vec{\mu}$, in which 
 case the energy scale of R parity violation 
 does not have to  be very small,  and therefore $\rho_{1,2,3}$
can take O(1) values. As a result,
the neutralinos and neutrinos can have a large mixing. 
The other possibility does not require the precise
alignment between $\vec{\rho}$ and $\vec{\mu}$, but the scale 
R-parity violation has to be  small compared to the weak scale. 
In this case all of the elements of ${\cal M}{\cal M}_0^{-1}$ is smaller than
one, and consequently the neutrino mass matrix can be obtained from the seesaw 
formula  $m_\nu={\cal M}{\cal M}_0^{-1}{\cal M}^T$.

Let us examine each case in more detail.
In our models discussed in the previous sections
( see the superpotential (\ref{superp2})),
we have $\mu_1=0$. The smallest non-zero eigenvalue $m_{\nu_3}$ of 
the mass matrix $M$ in the first case can be approximately written 
as \cite{ratm}
\be
m_{\nu_3} \simeq
\frac{M_Z^2(c_w^2 M_1+s_w^2 M_2)}{\vec{\mu}^2 M_1 M_2}~
[~\vec{\mu}^2~\vec{\rho}^2-(\vec{\mu}\cdot\vec{\rho})^2~]~,
\label{munu}
\ee
where
\be
\vec{\mu}=(\mu_0,\mu_1,\mu_2,\mu_3)~,
~\vec{\rho}=(\rho_0,\rho_1,\rho_2,\rho_3)~.
\label{vecmu}
\ee
Note that $\vec{\mu}$ and $\vec{\rho}$ do not contain
$\tilde{\mu}$ and $\rho_{\tilde{u}}, \rho_{\tilde{d}}$,
respectively. Using the angle $\xi$ made by
$\vec{\mu}$ and $\vec{\rho}$ and the GUT inspired relation
$M_1/M_2=(5/3)\tan^2\theta_W$, the neutrino mass (\ref{munu}) can
be written as
\be
m_{\nu_3} \simeq \frac{8}{5}\frac{M_W^2}{M_2}
\frac{ \sin^2\xi}{1+\tan^2\beta}~,
\ee
where we have defined $|{\vec \rho}|=\cos\beta$ which 
would coincide with $v_d/(v_u^2+v_d^2)^{1/2}$ 
of the R-parity conserving case
if only $H_u$ and $H_d$ would acquire a non-vanishing VEV.
To obtain a neutrino mass such as $\lsim  2.8$ eV
satisfying the combined mass bound coming
from the tritium $\beta$-decay \cite{tri} and various observations of the 
neutrino oscillation \cite{sno}, we need 
$\sin\xi \lsim 3\times 10^{-4}$ for $M_2= 1$ TeV and $\tan\beta=10$.
It may be interesting to see how the eigenstate $\psi_{\nu_3}$ of the 
smallest non-vanishing mass $m_{\nu_3}$ is composed.
Here we consider only the case in which $\psi_1$ and $\psi_2$ are decoupled
(that is, $\mu_1=\mu_2=0$).
Since ${\vec \rho}$ has to be  almost parallel to ${\vec \mu}$,
we make an approximation that  ${\vec \rho} \propto {\vec \mu}$, and 
find that the mass eigenstate is given by
\be
\psi_{\nu_3} &\simeq &\frac{1}{\sqrt{\mu_0^2+\mu_3^2}}
(\mu_0 \psi_3-\mu_3 \psi_d)~.
\ee
So the mixing between $ \psi_3 $ and $\psi_d $ will be large in general,
but no mixing with the other neutralinos.
There are two zero mass 
eigenvalues at tree level, but 
in higher orders
in perturbation theory \cite{loop,ratm}
this degeneracy is removed and 
the mixing among the neutrinos 
Although the couplings in the superpotential (\ref{superp3})
are restricted by a discrete R-symmetry (see Eq. (\ref{yukawas})),
three neutrinos mix  at one-loop order, allowing
a variety of mixing among neutrinos depending 
on the size of the R-parity violating
parameters. However, we cannot say more about its nature at present.

In the second case the neutrino mass matrix can be obtained from
the seesaw formula
\begin{equation}
m_\nu={\cal M}{\cal M}_0^{-1}{\cal M}^T=
{M_Z^2(c_w^2M_1+s_w^2M_2)\over M_1M_2\mu_0^2}
\left(\begin{array}{ccc}\Gamma_e^2 & \Gamma_e\Gamma_\mu 
&\Gamma_e\Gamma_\tau \\
 \Gamma_e\Gamma_\mu & \Gamma_\mu^2 
&\Gamma_\mu\Gamma_\tau \\
\Gamma_e\Gamma_\tau & \Gamma_\mu\Gamma_\tau 
&\Gamma_\tau^2 \\  \end{array}\right), 
\label{nmass}
\end{equation}
where $\Gamma_\alpha=-\rho_\alpha\mu_0 +\rho_0\mu_\alpha$.
The non-zero eigenvalue of this matrix is given by
\be
M_Z^2(c_w^2M_1+s_w^2M_2)\vert \vec{\Gamma}\vert^2\over M_1M_2\mu_0^2 \nn
\ee
which is equivalent 
to Eq.~(\ref{munu}) up to the higher order terms of $\mu_\alpha$ 
and $\rho_\alpha$.
A possible diagonalization matrix of (\ref{nmass}) 
 is \footnote{We assume
 that all the elements of $m_\nu$
 are real, and $V_{\nu}^{T} m_\nu V_{\nu}=$ diagonal.}
\be
V_\nu=\left(\begin{array}{ccc} \cos\gamma & \sin\gamma& 0 \\
-\sin\gamma & \cos\gamma &0 \\ 0 & 0 & 1 \\
\end{array}\right)
\left(\begin{array}{ccc} \cos\delta & 0 & \sin\delta \\
0  & 1 &0  \\ -\sin\delta &0 & \cos\delta \\
\end{array}\right)
\left(\begin{array}{ccc} 
\cos\alpha & \sin\alpha & 0 \\
  -\sin\alpha & \cos\alpha &0  \\ 0 &0 & 1 \\
\end{array}\right)~,
\ee
where $\tan\gamma=-\Gamma_\mu /\Gamma_e$,
$\tan\delta=\sqrt{\Gamma_e^2+\Gamma_\mu^2}/\Gamma_\tau$,
and 
$\alpha$ is an arbitrary angle. This arbitrariness results
from the fact that the mass matrix (\ref{nmass}) 
has two degenerate eigenvalues. 
Now to find the mixing matrix in the lepton sector $V^{\rm MNS}$, we remind 
ourselves that our R-charge assignment (see Table 2)
constrains the mass matrix of the charged leptons to have the form
\footnote{If we take other R-charge assignment for the lepton
sector, this feature cannot be realized.}
\begin{equation}
m_\ell=\left(\begin{array}{ccc} m_{ee} & 0 & 0 \\
0 & m_{\mu\mu} & m_{\mu\tau}  \\  0 &m_{\tau\mu}  & m_{\tau\tau} \\
\end{array}\right).
\end{equation}
This matrix can allow a maximum mixing in the $e$ and $\mu$ 
sector, which is favored for the realization of a bi-maximal
mixing in the lepton sector \cite{haba}.
(The  bi-maximal
mixing is considered to be a favored form to explain the solar 
and atmospheric neutrino observation.) 
Since  the mixing matrix  $V^{\rm MNS}$ is given by
 $V^{\rm MNS}=V_\ell^\dagger V_\nu$ ($V_\ell$ is the
 diagonalization matrix of the matrix $m_\ell$), the bi-maximal
 mixing form
 \be
V^{\rm MNS}\simeq \left(\begin{array}{ccc}
{1\over\sqrt 2} & {1\over\sqrt 2} & 0 \\ -{1\over 2} &{1\over 2} &
 -{1\over\sqrt 2} \\ -{1\over 2} &{1\over 2} & {1\over\sqrt 2} \\
\end{array}\right)
\ee
may be obtained if, for instance,
$\sin\delta\sim 0$ and 
$\cos(\alpha+\gamma)\sim\sin(\alpha+\gamma)\sim 1/\sqrt 2 $.
Note that the higher order corrections  resolve the
mass degeneracy and hence fix
the size of the angle $\alpha$,  and so we will need  more detailed study 
for a definite conclusion. We however expect to obtain
 results that are similar to those  in \cite{ratm},
in which, as far as the neutrino-neutralino sector is concerned,
similar models have been studied.

\subsection{The lightest Higgs}
 Since there exist two pairs of Higgs doublets in our models,
there exist four $CP$-even  neutral,
 three  $CP$-odd  neutral and three pairs of charged Higgs bosons
 that are mixed with the neutral and
 charged scalar leptons, respectively. 
 Here we are interested in the neutral sector, because
 we would like to find out the upper bound of
 the mass of the lightest Higgs boson.
 We denote the neutral scalar components of $H_u$ and $\tilde{H}_{u}$ by
$ h_{ui}$ with $i=1,2$, and those of the down-type ones
 $H_d, \tilde{H}_{d}$ and $L_i~(i=1,2,3)$
 by $ h_{d\alpha}$ with $\alpha=1,\dots,5$,
respectively. Then the most general renormalizable scalar potential
including the SSB terms can be written as
\be
V_N &=&(m^2_u)_{ij}h^{*}_{ui} h_{uj}+(m^2_d)_{\alpha\beta}
h^{*}_{d\alpha}h_{d\beta}
-(B_{i\alpha}  h_{ui}h_{d\alpha}+h.c.)\nn\\
& &+\frac{1}{8}(g^2+g^{\prime 2})(~
h^{*}_{ui} h_{ui}-h^{*}_{d\alpha}h_{d\alpha}~)^2~.
\label{scalarp}
\ee

Since physics is independent of the choice of a basis
of the fields, we go to a basis, in which
 only $h_{u1}$ and $h_{d1}$ have a non-vanishing
VEV. Accordingly we define
\be
h_{u1} &=&\frac{1}{\sqrt{2}}(v_u+\varphi_1+i\eta_1)~,~
h_{d1} =\frac{1}{\sqrt{2}}(v_d+\varphi_2+i\eta_2)~,\nn\\
h_{u2} &=&\frac{1}{\sqrt{2}}(\varphi_3+i\eta_3)~,
~h_{di} =\frac{1}{\sqrt{2}}(\varphi_{i+2}+i\eta_{i+2})~,~i=2,\dots,5~,
\label{varphi}
\ee
where $\varphi$'s and $\eta$'s are real scalar and pseudo-scalar
components of the Higgs fields, respectively.
In this basis, the mass matrices 
$M_E^2$ and $M_O^2$ for the
$CP$-even and $CP$-odd scalars, respectively, take the form
\be
{\bf M}^2_{E,O} & = &\left(
\begin{array}{ll} 
{\bf M}^{SM}_{E,O} &  {\bf B}_{E,O}  \\
 {\bf B}_{E,O}^{T} &  {\bf m}_{E,O}
\end{array} \right)~,
\label{even-odd}
\ee
where
\be
{\bf M}^{SM}_{E} &=&\left(
\begin{array}{ll} 
(v_d/v_u) B_{11} +\frac{1}{4}(g^2+g^{\prime 2})
v_u^2 &- B_{11}-\frac{1}{4}(g^2+g^{\prime 2})v_u v_d\\
- B_{11}-\frac{1}{4}(g^2+g^{\prime 2})v_u v_d &
  (v_u/v_d) B_{11}+\frac{1}{4}(g^2+g^{\prime 2})v_d^2
\end{array} \right)~,
\label{msm}\\
{\bf M}^{SM}_{O} &=&\left(
\begin{array}{ll} 
  (v_d/v_u)B_{11}  & B_{11}\\
 B_{11}& (v_u/v_d) B_{11} 
\end{array} \right)~,\\
{\bf B}_{E(O)} &=&\left(
\begin{array}{ll} 
(v_d/v_u)B_{12} &-(+) B_{1j} \\
  -(+)B_{12} &(v_u/v_d) B_{1j} 
\end{array} \right)~,~j=2,\dots,5~,\\
({\bf m}_{E})_{3 3} & =&({\bf m}_{O})_{3 3}= ( m_u^2)_{22} +
\frac{1}{8}(g^2+g^{\prime 2})(v_u^2-v_d^2)
~,\\
({\bf m}_{E(O)})_{3 j+2} &=&({\bf m}_{E(O)})_{j+23}= 
 -(+) B_{2j} ~,~j=2,\dots,5~,\nn \\
({\bf m}_{E})_{i+2 j+2} & =& 
({\bf m}_{O})_{i+2 j+2}=(m_d^2)_{ij}+
 \frac{1}{8}(g^2+g^{\prime 2})(v_d^2-v_u^2)\delta_{ij}
 ~,~i,j=2,\dots,5~.
 \label{m5}
\ee
To derive the above formulas we have used minimum conditions of the
scalar potential and also assumed that all the parameters 
appearing in the scalar potential (\ref{scalarp}) 
are real \footnote{The mass parameters above
are not those defined in the original scalar potential (\ref{scalarp}).
They correspond to those in the new basis in which 
only $h_{u1}$ and $h_{d1}$ acquire a 
non-vanishing VEV.}.
Note that the upper $2\times 2$ matrices $ {\bf M}^2_{E} $  and
$ {\bf M}^2_{O} $ 
have exactly the same form as those of the MSSM.
We see from Eqs. (\ref{even-odd})
--(\ref{m5}), as in the case of the MSSM,  that
$\mbox{Tr}~ {\bf M^2_E}= M_Z^2+\mbox{Tr}~ {\bf M^2_O}$
is satisfied, which yields the  sum rule at the tree level 
\be
m_h^2+\sum_{i=1}^{3}m_{Hi}^2+\sum_{i=1}^{3}m_{\tilde{\nu}_+ i}^2
&=&M_Z^2+\sum_{i=1}^{3} m_{Ai}^2+
\sum_{i=1}^{3}m_{\tilde{\nu}_- i}~,
\ee
where $m_h, m_{H} and m_{\tilde{\nu}_+}$ stand for the masses of the
$CP$-even scalars, and $m_{A}, m_{\tilde{\nu}_-}$ for the 
$CP$-odd scalars. 

Now we come to discuss  the lightest Higgs mass $m_h$.
To this end, we concentrate on the size of
the diagonal elements of ${\bf M}^2_{E}$ and ${\bf M}^2_{O}$,
because their smallest eigenvalues cannot be larger than the
smallest  diagonal elements.
The scalar mass-squared $(m^2_u)$ and $(m^2_d)$
in the scalar potential (\ref{scalarp})
consist of both the SSB scalar mass-squared and the contribution
from the superpotential (\ref{superp2}). Here we remind ourselves
that all the parameters belonging to  the mass as well as interaction
terms that involve at least one
of $\tilde{H}_{u}$ or $\tilde{H}_{d}$ are 
very much suppressed at the escape energy. In particular,
the SSB scalar mass-squared for 
$\tilde{H}_{u}$ or $\tilde{H}_{d}$ (which we denote by
 $(\tilde{m}^2_u)_{22}$ and $(\tilde{m}^2_d)_{22}$)
 vanishes at the superconformal
 fixed point \cite{karch, luty, nelson},
 if the weakly coupled low-energy sector is switched off.
 It has been however found in \cite{kobayashi1} that the low-energy
 sector has a non-trivial influence on
 their evolution such that
 they rather approach, translated into the present case, as
 \be
(\tilde{m}^2_u)_{22} &\simeq & (\tilde{m}^2_d)_{22} \to 
(\gamma^*)^{-1} \frac{ 3g^2}{8 \pi^2} |M_2|^2~,
 \ee
 where $\gamma*$ is the anomalous
 dimension of $\tilde{H}_{u}$ (or  $\tilde{H}_{d}$) at the fixed point 
 (see Eq. (\ref{gammastar})),  $M_2$ is 
 the $SU(2)_L$  gaugino mass, and
 we have neglected  the $U(1)_Y$ contribution.
Below the escape energy $\Lambda_C$, their
evolution is dictated by the low-energy sector, and
all the couplings
that contribute to the evolution,
 except for the gauge couplings of this sector,
are suppressed because of the superconformal force.
>From this consideration we obtain approximately
$(\tilde{m}^2_u)_{22}$ and $(\tilde{m}^2_d)_{22}$ at $M_Z$
\be
 (\tilde{m}^2_u)_{22} &\simeq&
 (\tilde{m}^2_d)_{22} \simeq \frac{3g^2}{8\pi^2}|M_2|^2
~\Big[~(\gamma^*)^{-1}+ \ln \frac{\Lambda_C}{M_Z}~\Big]~,
\ee
where the quantity in the parenthesis is a positive number
of $ \gsim O(1)$.
Consequently, the total contributions to the 
diagonal elements in question can be written as
\be
 (m^2_u)_{22} = ({\vec \mu})^2+ (\tilde{m}^2_u)_{22}+
 \frac{1}{2}M_Z^2 \cos 2\beta~, \quad
  (m^2_d)_{22} = \tilde{\mu}^2+ (\tilde{m}^2_d)_{22}+
 \frac{1}{2}M_Z^2 \cos 2\beta~,\nn
\ee
where  $\tan\beta =v_u/v_d$ is defined in
the basis in which all the VEVs 
except for $H_{u1}$ and $H_{d1}$ vanish
(see  Eq. (\ref{varphi}), and
 ${\vec \mu}$ is given in Eq. (\ref{vecmu}).).
It is then obvious
that we  can make   $(m^2_u)_{22}$ and $ (m^2_d)_{22}$ 
arbitrarily large by making the gaugino mass $M_2$ large.
Therefore,  the smallest eigenvalue of ${\bf M}^{2}_{E}$ sits
in ${\bf M}^{SM}_{E}$, implying that
we have the same upper bound of the lightest Higgs
as in the case of the MSSM
\be
 m_h^2 &\leq & M_Z^2 \cos^2 2\beta~,
\label{bound}
 \ee
because the matrix ${\bf M}^{SM}_{E}$ ( given in Eq. (\ref{msm}) ) has exactly
the same form as in the MSSM.
The tree-level bound  (\ref{bound}) should be of course corrected
in higher orders in perturbation theory \cite{higgsmass,higgs}.
We expect that the correction will be very similar to the case
of the MSSM, especially if the other masses  are large.

\section{Conclusion}
In supersymmetric standard models
with  R-parity and lepton number
violations,  the left-handed lepton and  down-type Higgs supermultiplets
should be treated on the same footing, unless there exist further
quantum numbers that distinguish them from each other.
Therefore, the $\mu$ problem in these models is closely related
to the question of why the neutrinos are so light.
In this paper we have proposed to solve 
the $\mu$ problem  in this class of models
by coupling the models to  a superconformal gauge force.
We found that for this idea to work we have to extend
the MSSM so as to contain at least another pair of
Higgs doublets, which mediate the superconformal suppression
to the MSSM sector. 
We have shown that
a suppression of 
$ \lsim O(10^{-13})$ for the $\mu$ parameter and
neutrino masses can be  achieved generically.

We have constrained the form of the superpotential 
of the low energy sector by
imposing an anomaly-free discrete R-symmetry, while we
have allowed most general, renormalizable  supersymmetry breaking
terms. We have found that the discrete R-symmetry
automatically suppress
 the lepton-flavor
violating processes such as $\mu\rightarrow e~\gamma,
~\mu\rightarrow e~e~e,$ $\mu$-$e$ conversion in nuclei, the electron 
EDM and  also 
the neutrinoless double $\beta$ decay.
The resulting models can accommodate to
a large mixing among neutrinos,
and it has turned out that  the upper bound
of the lightest Higgs mass  
of the MSSM remains unchanged in these
extended models.
Finally we expect
that the escape energy of the superconformal sector is 
$\lsim$ O(10) TeV so that this sector could be
experimentally observed in near feature.

\vspace{0.5cm}
\noindent
{\large \bf Acknowledgments}\\
This work is supported partially by  the Ministry of
Education, Science 
and Culture and by
 the Japan Society for the Promotion of Science
 (No. 11640267 and 12047213).
We would like to thank H. Terao
for useful discussions.

\end{document}